\pdfoutput=1
\documentclass[fleqn,usenatbib]{mnras}

\usepackage{newtxtext,newtxmath}

\usepackage[T1]{fontenc}

\DeclareRobustCommand{\VAN}[3]{#2}
\let\VANthebibliography\thebibliography
\def\thebibliography{\DeclareRobustCommand{\VAN}[3]{##3}\VANthebibliography}

\usepackage{graphicx}	
\usepackage{amsmath}	
\usepackage{multirow}
\usepackage{booktabs}
\usepackage{caption}
\usepackage{subcaption}
\usepackage{xcolor}

\title[Chandra source classification using ML]{Automated classification of {\em Chandra} X-ray point sources using machine learning methods}

\author[Kumaran et al.]{
Shivam Kumaran$^{1}$\thanks{E-mail: kumaranshivam57@gmail.com},
Samir Mandal$^{1}$,
Sudip Bhattacharyya$^{2}$,
Deepak Mishra$^{3}$
\\
$^{1}$Department of Earth and Space Sciences, Indian Institute of Space Science and Technology, Thiruvananthapuram, 695547, India\\
$^{2}$Department of Astronomy and Astrophysics, Tata Institute of Fundamental Research, Mumbai, 400005, India\\
$^{3}$Department of Avionics, Indian Institute of Space Science and Technology, Thiruvananthapuram, 695547, India
}

\date{Accepted 2023 January 31. Received 2022 December 28; in original
form 2022 September 14}

\pubyear{2023}

\begin{document}
\label{firstpage}
\pagerange{\pageref{firstpage}--\pageref{lastpage}}
\maketitle

\begin{abstract}
A large number of unidentified sources found by astronomical surveys and other observations necessitate the use of an automated classification technique based on machine learning methods. The aim of this paper is to find a suitable automated classifier to identify the point X-ray sources in the {\em Chandra} Source Catalogue (CSC) 2.0 in the categories of active galactic nuclei (AGN), X-ray emitting stars, young stellar objects (YSOs), high-mass X-ray binaries (HMXBs), low-mass X-ray binaries (LMXBs), ultra luminous X-ray sources (ULXs), cataclysmic variables (CVs), and pulsars. The catalogue consists of $\approx 3,17,000$ sources, out of which we select 2,77,069 point sources based on the quality flags available in CSC 2.0. In order to identify unknown sources of CSC 2.0, we use multi-wavelength features, such as magnitudes in optical/UV bands from {\em Gaia}-EDR3, {\em SDSS} and {\em GALEX}, and magnitudes in IR bands from {\em 2MASS}, {\em WISE} and {\em MIPS-Spitzer}, in addition to X-ray features (flux and variability) from CSC 2.0. We find the Light Gradient Boosted Machine, an advanced decision tree-based machine learning classification algorithm, suitable for our purpose and achieve 93\% precision, 93\% recall score and 0.91 Mathew’s Correlation coefficient score. With the trained classifier, we identified 54,770 (14,066) sources with more than $3\sigma$ ($4\sigma$) confidence, out of which there are 32,600 (8,574) AGNs, 16,148 (5,166) stars, 5,184 (208) YSOs, 439 (46) HMXBs, 197 (71) LMXBs, 50 (0) ULXs,  89 (1) CVs,  and 63 (0) pulsars. This method can also be useful for identifying sources of other catalogues reliably.

\end{abstract}




\section{Introduction}

Huge amounts of high-quality data from many astronomical sources are becoming available due to large-scale surveys and an open data access policy.
Many of these sources are unidentified. 
These data have a great potential for discoveries of novel classes of sources, new sources of known classes, new observational phenomena and even perhaps new physics.
However, the sheer volume of data  necessitates taking an automated approach to source classification. Such an automated classifier can be designed using Machine Learning (ML) methods \citep{2010IJMPD..19.1049B}. ML algorithms are capable of learning patterns in big data and can identify decision boundaries based on the already identified examples. Unlike the manual methods of identifying sources, for example, based on the colour-colour diagram clustering limited to 3-dimension, ML methods can create decision boundaries in very high dimensional feature space. 

In optical/IR astronomy, several works have been done for source identification using machine learning \citep{2016A&A...596A..39K, 2020MNRAS.495.4135T, 2019MNRAS.488.3358T, 2021MNRAS.506..677C}. 
However, in X-ray astronomy, the use of machine learning is relatively less.
\cite{2015ApJ...813...28F}  did classification of variable sources in the third {\em XMM-Newton} Serendipitous Source Catalogue (3XMM) using a Random Forest classifier with timing properties. \citet{2021MNRAS.503.5263Z} used Random Forest and LogiBoost to classify the sources in {\em XMM-newton's}  4XMM-DR9 using multiwavelength properties from  {\em GAIA}, {\em WISE}, and {\em 2MASS}. Classification of X-ray binaries based on whether the compact object is a black hole or a neutron star was done by \citet{2022arXiv220400346D} using {\em MAXI/GSC} lightcurve. \citet{2022MNRAS.510..161F} used Random Forest and AdaBoost to develop an automated classifier for the identification of AGN and to classify them as Type-I or Type-II AGN further with the data from {\em XMM-Newton} and SDSS. \citet{2022A&A...657A.138T} used multiwavelength data to classify sources in {\em Swift-XRT} and {\em XMM-Newton} serendipitous Source Catalogues using Naive Bayes classifier. \cite{2021MNRAS.501.3457P} used the spectrum in the energy range of 5-25 keV from the {\em Rossi X-ray Timing Explorer (RXTE)} to identify the nature of the compact object in the Low Mas X-ray Binary objects using Random Forest classifier.

In this paper, we aim to classify the point X-ray sources in the {\em Chandra} Source Catalogue (CSC) 2.0, which is the second source catalogue \citep{2020AAS...23515405E} of the {\em Chandra X-ray Observatory}. A unique strength of {\em Chandra} is its angular resolution ($\leq 1^{\prime \prime}$), which is substantially better than the FWHM ($6^{\arcsec}$; \cite{2001A&A...365L...1J}) of {\em XMM-NEWTON} and on-axis angular resolution (5$^{\arcsec}$; \cite{1982AdSpR...2d.251A}) of {\em ROSAT}. CSC 2.0 contains properties of about 3,17,000 sources from the observations till the end of 2014 and total sky coverage of $558.65$ deg$^2$. Most of these sources remain unidentified. 

To the best of our knowledge, an approach adopted in this paper on
automated classification of unidentified sources in CSC 2.0 has
not been published yet. Due to the sub-arcsec spatial resolution and high sensitivity of \textit{Chandra}, the source spatial population density in the CSC 2.0 is the highest among X-ray source catalogues. Thus, CSC 2.0 offers an excellent opportunity for the serendipitous discovery of objects of known classes and new exotic objects \citep{2019HEAD...1710929M}. 

In this paper, we discuss the development of an automated classifier based on supervised machine-learning algorithms for the point sources in CSC 2.0. The classifier primarily uses the features available in CSC 2.0, which are flux in five different bands of {\em Chandra}'s ACIS instrument and variability properties. In addition to the X-ray features, the source identification can be improved with the use of features available in other wavelengths. We obtain multiwavelength features from {\em 2MASS}, {\em WISE}, {\em Gaia}-EDR3, {\em MIPS-Spitzer}, {\em SDSS} and {\em GALEX}. We explore decision tree based supervised machine learning classification algorithms. We find that the Light Gradient Boosted Machine gives the best classification performance. 

In the \S \ref{sec:data-source}, we describe the details of the data, standardizing the data and the method for identifying the training set. In \S \ref{sec:method}, we present various classifier models which we explore and the methodology for selection and validation of the classifier. In \S \ref{sec:result}, we present the result of the model validation and performance evolution. 
In \S \ref{sec:Summary}, we give a summary and conclusions.

\section{The Data}\label{sec:data-source}
The objective of a machine learning (ML) classification model is to learn the relation between the features of a sample and its class label. In supervised ML methods, this relation is learnt using already labelled samples. For astrophysical objects, the features can be observed properties like magnitudes or flux in various wavelength bands. CSC 2.0 provides tabulated values of various  observed properties. Apart from X-ray features, the source properties in other multiwavelength (MW) bands can be used to improve the classification further.

\subsection{X-ray Data}\label{sec:x-ray-data}
{\em Chandra} have two focal plane instruments: Advanced CCD Imaging Spectrometer ({\em ACIS}) and High Resolution Camera ({\em HRC}). The {\em ACIS} instrument observes in broad (b): 0.5-7.0 keV, ultrasoft (u): 0.2-0.5 keV, soft (s): 0.5-1.2 keV, medium (m): 1.2-2.0 keV, and hard (h): 2.0-7.0 keV bands. The {\em HRC} instrument observes in 0.1-10 keV energy band and is designated as `W' band. 
CSC 2.0 was prepared with the observations from {\em ACIS} and {\em HRC} till the end of 2014. It contains the information of 3,17,167 sources, out of which 2,96,473 are point sources, selected by the parameter $extent\_flag==0$ in the catalogue. We further filter the sources based on the quality flags available in CSC 2.0, which are: pileup\_flag, sat\_src\_flag, conf\_flag, streak\_src\_flag (Table \ref{tab:flag-table}).
\begin{table}
	\centering
	\caption{Quality flags used to filter sources in CSC 2.0 and their description.}
	\label{tab:flag-table}
        \begin{tabular}{lr}
        \hline
        \textbf{Flag code}         & \textbf{Description}                                            \\ \hline
        pileup\_flag      & ACIS pile-up fraction exceeds $\sim$10\%  \\
        sat\_src\_flag    & Saturated source in all observations \\
        conf\_flag        & Source confused (source and/or background \\ 
                          & regions in different stacks may overlap) \\
        streak\_src\_flag & Source located on ACIS CCD read-out streak                  \\ \hline
        \end{tabular}
\end{table}

 For CSC 2.0, the energy flux in each band is determined using aperture photometry. The source count is derived from an elliptical source region and subtracted by the background count in the surrounding region. To convert the count rate to energy flux, the total count rate is summed up and then scaled by the local ancillary response function. In this work, we use aperture-corrected average net-flux in b, u, s, m, and h bands, which are named as b-csc, u-csc, s-csc, m-csc, h-csc (Table \ref{tab:feature-table}) respectively. 

 The CSC also include various features indicating the short-term temporal variability (intra-observation) and long-term temporal variability (inter-observation) in each energy band. 

The intra-observation variability features are calculated using three methods:
 \begin{enumerate}
     \item Gregory-Loredo variability probability

     \item Kolmogorov-Smirnov test

     \item Kuiper's test 

 \end{enumerate}

The Gregory-Loredo (GL) \citep{1992ApJ...398..146G} analysis provides the probability ({\em var\_prob}) that the count rate in the source region varies over a flat distribution for each observation. 
It also calculates the odds ratio, time-integrated average count rate and standard deviation ($\sigma$). The inverse of the odds ratio represents the significance of the observed distribution. GL algorithm defines a variability index {\em var\_index} (an integer between 0-10)\footnote{See variability index criterion table: \url{https://cxc.cfa.harvard.edu/csc/why/gregory\_loredo.html\#varindex}} using {\em var\_prob}, odds ratio and the fraction of lightcurve within $3\sigma$ and $5\sigma$. Any source having {\em var\_index} > 6 is considered a variable. The highest {\em var\_prob} and {\em var\_index} across the observations are recorded as {\em var\_intra\_prob} and {\em var\_intra\_index} in the master table for the given source.

 Kolmogorov-Smirnov (K-S) test calculates the probability that the arrival times of the events within the source region are inconsistent with a constant source count rate. It compares the cumulative photon distribution in the source region with a constant count rate. 

 The highest probability value across the observations is recorded as {\em ks\_intra\_prob} in the master table for the given source. Similarly, a variability index, {\em kp\_intra\_index}, is defined using Kuiper's test. While the K-S test is sensitive around the median value, Kuiper's test is more effective towards the tail of a probability distribution.
 
 The inter-observations variability probability ({\em var\_inter\_prob}) 
 defines the variation of the photon flux between the contributing observations. The maximum log-likelihood functions of the observed photon flux distribution are calculated assuming a different flux for each observation and constant photon flux. The difference between these two maximum log-likelihood functions follows a $\chi^2$ distribution. The {\em var\_inter\_prob} represents the cumulative probability of this $\chi^2$ distribution. The inter-observation variability index, {\em var\_inter\_index} (an integer between 0-8), is defined based on {\em var\_inter\_prob} and the number of degrees of freedom. The variance of the individual observation fluxes is indicated by the feature {\em var\_inter\_sigma} parameter.

 In this work, we use four intra-observation variability properties in b-band ({\em var\_intra\_prob\_b, var\_intra\_index\_b, ks\_intra\_prob\_b, kp\_intra\_prob\_b)} and three inter-observation variability properties in b-band ({\em var\_inter\_prob\_b, var\_inter\_index\_b and var\_inter\_sigma\_b}).
  
 We use Chandra Interactive Analysis of Observations (CIAO-4.14) \citep{2006SPIE.6270E..1VF} to download the data from CSC 2.0 using Astronomical Data Query Language (ADQL)\footnote{\url{https://www.ivoa.net/documents/ADQL/20180112/PR-ADQL-2.1-20180112.html}}. 

\subsection{Multiwavelength (MW) Data}\label{sec:mw-data}

AllWISE catalogue \citep{2014yCat.2328....0C} is the all infrared survey catalogue built by combining the data from {\em Wide-field Infrared Survey Explorer (WISE)} mission's two all-sky survey projects: {\em WISE} cryogenic phase \citep{2010AJ....140.1868W} and the post cryogenic NEOWISE survey \citep{2011ApJ...731...53M}. The AllWISE  is an all-sky infra-red survey at the wavelength bands 3.4 $\mu$m, 4.6 $\mu$m, 12 $\mu$m, and 22 $\mu$m named W1, W2, W3 and W3 bands, respectively. The AllWISE catalogue contains 747,634,026 sources with limiting sensitivities W1 < 17.1, W2 < 15.7, W3 < 11.5 and W4 < 7.7 magnitude. In our work, we use W1, W2, W3 and W4 magnitude from the AllWISE catalogue.

The {\em Gaia} \citep{2016A&A...595A...1G} is an optical telescope launched and operated by European Space Agency. The {\em Gaia} Early Data Release-3 ({\em Gaia}-EDR3) \citep{2021A&A...649E...1F} contains 1,811,709,771 sources and gives magnitudes in three broadband optical passbands, green ($G$), blue ($G_{BP}$) and red ($G_{RP}$) passbands. In this work, we use Gaia-EDR3 $G$, $G_{BP}$ and $G_{RP}$ band magnitudes. We find the association with \textit{Gaia} using CDS X-match positional cross-match service \citep{boch2014cds} such that the source must be within $3\sigma$ positional error of CSC 2.0 and \textit{Gaia} EDR-3 error circle.

The 2 Micron All Sky Survey (2MASS) \citep{2006AJ....131.1163S} is the survey of the entire celestial sphere with a 99.99\% sky coverage in the infra-red domain at $1.25\mu$m (J), 1.65$\mu$m (H) and 2.16$\mu$m (K\textsubscript{s}) bands. The survey data were taken by two identical telescopes of diameter 1.3 m at Arizona and Chile in the northern and southern hemispheres, respectively. The Survey catalogue contains 470,992,970 point sources.

     \begin{table*}
    \centering
    \thispagestyle{empty}
    \caption{Multi-wavelength features from various catalogues used in this work.}
    \label{tab:feature-table}
    \begin{tabular}{lcr}
    \hline
    \textbf{Feature Source} & \textbf{Feature Name}        & \textbf{Feature Description}                                      \\ \hline
    CSC 2.0              & gal\_l2              & Galactic longitude                                        \\
                     & gal\_b2              & Galactic Latitude                                         \\
                     & b-csc                & Flux in ACIS broad (b) band (0.5-7.0 keV)                 \\ 
                     & u-csc                & Flux in ACIS ultrasoft (u) band (0.2-0.5 keV)             \\
                     & s-csc                & Flux in ACIS soft (s) band (0.5-1.2 keV)                  \\
                     & m-csc                & Flux in ACIS medium (m) band (1.2-2.0 keV)                \\
                     & h-csc                & Flux in ACIS hard (h) band (2.0-7.0 keV)                  \\
         & var\_inter\_prob\_b  & Inter-observation variability probability in ACIS b band            \\
         & var\_inter\_sigma\_b & Standard deviation in Inter-observation flux variability            \\
                     & var\_inter\_index\_b & Inter-observation variability index                       \\
         & var\_intra\_prob\_b  & Intra-observation Gregory-Loredo variability  probability in b band \\
         & ks\_intra\_prob      & Kolmogorov-Smirnov Intra-observation variability probability b-band \\
         & kp\_intra\_prob\_b   & Intra-observation Kupier's test variability probability in b band   \\
                     & var\_intra\_index\_b & Intra-observation variability index                       \\ 
    GAIA-EDR3             & G                    & Gaia Green (G) pass-band magnitude                        \\
                     & Bp                   & Gaia Blue (G\_BP) pass-band magnitude                     \\
                     & Rp                   & Gaia Red (G\_RP) pass-band magnitude                      \\ 
                     
    GALEX            & FUV                  & Magnitude in GALEX FUV band                               \\
                     & NUV                  & Magnitude in GALEX NUV band                               \\ 
    SDSS             & u-sdss               & SDSS u band magnitude                                     \\
                     & g-sdss               & SDSS g band magnitude                                     \\
                     & r-sdss               & SDSS r band magnitude                                     \\
                     & i-sdss               & SDSS i band magnitude                                     \\
                     & z-sdss               & SDSS z band magnitude                                     \\ 
    WISE             & W1                   & WISE W1(3.4 micron) band magnitude                        \\
                     & W2                   & WISE W2(4.6 micron) band magnitude                        \\
                     & W3                   & WISE W3 (12 micron) band magnitude                        \\
                     & W4                   & WISE W4 (22 micron) band magnitude                        \\ 
    {\em MIPS-Spitzer} & 24\_microns\_{\em MIPS}  & Magnitude in 24 micron band of {\em MIPS} on Spitzer                      \\ 
    2MASS            & J                    & J-band (1.235 micron) band magnitude                      \\
                     & H                    & H-band (1.662 micron) band magnitude                      \\
                     & K\_s                    & Ks-band (2.159 micron) band magnitude                     \\ 
    
    Colour$^*$ & B-R                  & Magnitude in Gaia Bp - magnitude in Gaia Rp               \\
                     & G-J                  & Magnitude in Gaia G band  - magnitude in 2MASS J band     \\
                     & G-W2                 & Magnitude in Gaia G band  - magnitude in WISE W2 band     \\
                     & Bp-H                 & Magnitude in Gaia G\_BP band  - magnitude in 2MASS H band \\
         & Bp-W3                & Magnitude in Gaia G\_BP band  - magnitude in WISE W3 band           \\
                     & Rp-K                 & Magnitude in Gaia G\_RP band  - magnitude in 2MASS K band \\
                     & J-H                  & Magnitude in 2MASS H - magnitude in 2MASS H band          \\
                     & J-W1                 & Magnitude in 2MASS J - magnitude in WISE W1 band          \\
                     & W1-W2                & Magnitude in WISE W1 - magnitude in WISE W2 band          \\ \hline
    
    \end{tabular}%
    \\
    $^*$The `colour' features are computed using available magnitude values.
    \end{table*}

{\em The Multiband Imaging Photometer (MIPS)} onboard {\em Spitzer} \citep{2004ApJS..154...25R, 2013AAS...22134006C} covers the infrared spectrum in the wavebands of 24 $\mu$m, 70 $\mu$m, and 160 $\mu$m. In this work, we use the 24$\mu$m band (bandwidth $\sim 5\mu$m) data due to its highest photometric accuracy of $6^{\arcsec}$.

The telescope {\em Galaxy Evolution Explorer  (GALEX)}  takes observation in the far ultraviolet (FUV: 1344–1786 \AA ) and near ultraviolet (NUV: 1771–2831 \AA) wavelengths with a resolution of $\sim 4.5^{\arcsec}$ (FWHM) and $\sim 6.0^{\arcsec}$ (FWHM) respectively \citep{2005ApJ...619L...7M}.

The Sloan Digital Sky Survey (SDSS; \cite{2000AJ....120.1579Y}) is an extensive photometric and spectroscopic survey with an astrometric accuracy of the order of $0.1\arcsec$. The SDSS obtains the data in five optical bands:u, g, r, i and z with the central wavelengths of 3560{\AA}, 4680{\AA}, 6180{\AA}, 7500{\AA}, and 8870{\AA} respectively. The limiting magnitudes of u, g, r, i and z are 21.6, 22.2, 22.2, 21.3, and 20.7, respectively. This work uses the 16\textsuperscript{th} data release of SDSS (SDSS-DR16; \cite{2020ApJS..249....3A}). This release includes the data from the previous release combined with the Apache Point Observatory Galactic Evolution Experiment 2 (APOGEE-2) survey and from the Extended Baryon Oscillation Spectroscopic Survey (eBOSS).

We use NASA/IPAC Extragalactic Database (NED) to obtain multi-wavelength information for the CSC sources. With the November 2021 release, NED integrated the CSC. With the help of the cross-match algorithm Match Expert \citep[MatchEx; ][]{2015ASPC..495...25O}, 80\% of the CSC sources were found to have associations with already existing objects, and 20\% became new objects in the NED database. We use CSC 2.0 names as the object identifier in NED to obtain the multiwavelength property with identifier-based query using \texttt{astroquery} package. We obtain multiwavelength data for 2,77,069 sources from NED. However, the NED server responded with error messages and the data could not be retrieved for 648 sources. Out of 2,77,069 objects in CSC 2.0, we could find an association for 60\% sources in {\em Gaia-EDR3}, 55\% in 2MASS, 43\% for \textit{MIPS-Spitzer}, 41\% for {\em WISE}, 24\% for {\em SDSS} and 17\% for {\em GALEX}.

The multiwavelength features used from these catalogues are given in Table \ref{tab:feature-table}. Besides these features, we compute the colours from the magnitude available in different bands and use them as additional features. We use an online multiwavelength visualization tool developed by \cite{2021RNAAS...5..102Y} to identify the colours that show the best class-wise clustering in a colour-colour diagram.

    \begin{table*}
        \centering
        \caption{Published catalogues used to identify sources in various classes.}
        \label{tab:source-cat-detail}
        \begin{tabular}{@{}lccr@{}}
        \toprule
        \textbf{Class} & \textbf{Catalogue source} & \textbf{Catalogue Details} & \textbf{Reference} \\ \midrule
        \textbf{AGN} & VERONCAT & Veron Catalogue of Quasars \& AGN, 13th Edition & \citep{veron2010catalogue} \\
       \textbf{STAR} & SKIFF & Catalogue of Stellar Spectral Classifications & \cite{2013yCat....102023S} \\
       \multirow{2}{*}{\textbf{YSO}} & & The Spitzer Space Telescope Survey ... & \cite{2012AJ....144..192M} \\
         &  & The Spitzer/IRAC Candidate YSO Catalogue ... & \cite{2021ApJS..254...33K} \\
        \textbf{HMXB} & HEASARC & SMCPSCXMM & \cite{sturm2013xmm} \\
         &  & High-Mass X-Ray Binaries Catalogue & \cite{liu2006catalogue} \\
         &  & INTEGRAL Reference Catalogue & \cite{ebisawa2003high} \\
         &  & Magellanic Clouds High-Mass X-Ray Binaries Catalogue & \cite{liu2005meg} \\
         &  & IBIS/ISGRI Soft Gamma-Ray Survey Catalogue  & \cite{hmxb-bird-2016}\\
         &  & INTEGRAL/ISGRI Catalogue of Variable X-Ray Sources & \cite{hmxb-tel-2010} \\
        \multirow{15}{*}{\textbf{LMXB}} & \multirow{15}{*}{HEASARC} & NGC 3115 Chandra X-Ray Point Source Catalogue & \cite{lmxb-ngc-3115} \\
         &  & Ritter Low-Mass X-Ray Binaries Catalogue  & \cite{lmxb-ritter-2003} \\
         &  & Low-Mass X-Ray Binaries Catalogue  & \cite{lmxb-liu-2007} \\
         &  & INTEGRAL Reference Catalogue  & \cite{ebisawa2003high}  \\
         &  & XMM-Newton M 31 Survey Catalogue  & \cite{lmxb-xmm-2005} \\
         &  & M 31 ... Point Source Catalogue & \cite{lmxb-m31-2013} \\
         &  & ROSAT All-Sky Survey & \cite{lmxb-rosat-2009} \\
         &  & INTEGRAL IBIS Hard X-Ray Survey & \cite{lmxb-integral-2015} \\
         &  & INTEGRAL IBIS 9-Year Galactic Hard X-Ray Survey Catalog& \cite{lmxb-integral-2012} \\
         &  & IBIS/ISGRI Soft Gamma-Ray Survey Catalogue & \cite{hmxb-bird-2016} \\
         &  & M 31 XMM-Newton ... X-Ray Point Source Catalogue & \cite{lmxb-xmm-2009} \\
          \textbf{ULX} & ULXRBCAT & & \cite{ulx-liu-2005}\\
         \textbf{CV} & The Open CV Cat. & The Open Cataclysmic Variable Catalogue  & \cite{2020RNAAS...4..219J} \\
        \multirow{2}{*}{\textbf{PULSAR}} & ATNF &  & \cite{2005AJ....129.1993M} \\
         & FERMI LAT (4FGL) & Fermi LAT Second Catalogue of Gamma-Ray Pulsars (2PC) & \cite{fermi-lat2013ApJS..208...17A} \\
         \bottomrule
        \end{tabular}%
    \end{table*}
    
 \subsection{The Training Set}
    We aim to classify the point sources in CSC 2.0 into the following classes: Active Galactic Nuclei (AGN), X-ray emitting stars (STAR), Young Stellar Objects (YSO), High Mass X-ray binaries (HMXB), Low Mass X-ray binaries (LMXB), Ultra Luminous X-ray sources (ULX), Cataclysmic Variables (CV) and pulsars. First, we prepare a list of already identified sources belonging to these classes. Various published catalogues that we use to identify known sources are given in Table \ref{tab:source-cat-detail}.
    
   We cross-match the coordinates of the known sources with all the 2,77,069 sources in our list. We use \textsc{Astropy}\footnote{This work made use of Astropy: \url{http://www.astropy.org}, a community-developed core Python package and an ecosystem of tools and resources for astronomy}, which is a \textsc{Python} package to perform cross-matching. We select a cross-match radius of $1^{\arcsec}$. In case there is more than one source in the cross-match radius, we consider the source with the least angular separation from the target source. Using this, we identify a total of 7,703 sources, of which there are 2395 AGNs, 2790 stars, 1149 YSOs, 748 HMXBs, 143 LMXBs, 211 ULXs, 166 CVs and 101 pulsars. The class-wise percentages of identified sources are given in Table \ref{tab:train-set}. These sources are used to train the supervised machine learning algorithm.
   In our training set, we have a large fraction of AGNs, stars, and YSOs, which comprise a total of about 80\% of the entire training set. The classes LMXB, ULX, CV and pulsar are minorities with populations of only 1-3\% of the training set. 
     
     We create a data-table with 41 MW features (Table \ref{tab:feature-table}) from CSC-2.0, {\em GAIA}-EDR3, {\em 2MASS}, {\em SDSS}, {\em WISE}, {\em GALEX}, and {\em MIPS-Spitzer} for 2,77,069 point sources in CSC 2.0. Out of this, we keep a separate data table of 7,703 known sources as the training data set. We attempt to identify the rest (2,69,366) of the sources.
    
\section{Methodology}\label{sec:method}
    We prepare the multi-wavelength data for a set of already identified sources. We use this set to train a supervised machine-learning model. The model learns the pattern of the features in the training set and identifies the best possible decision boundary in the feature space. For designing the machine learning classification model, we use the python package \texttt{Scikit-Learn} \citep{scikit-learn}. In \texttt{Scikit-Learn}, various ML models are defined as Python classes with several options to customize the model. These models also implement general routine functions like `fit' to train the model and use the trained model to `predict' the class of a new sample. From \texttt{Scikit-Learn}, we test Multi-Layer Perceptron, K-Nearest Neighbour, Random Forest and Gradient Boosted Decision Trees. 
    We find that decision tree-based models---Random Forest (RF) and Gradient Boosted Decision Tree (GBDT)---perform better than other models. We also explore the Light Gradient Boosted Machine (LightGBM) \citep{ke2017lightgbm}, which is an advanced development over GBDT, in this work. 
    \subsection{Classifier Models}
        \subsubsection{Random Forest}
        Random Forest (RF; \cite{randomforest}) is an ensemble of decision trees. Each decision tree is built from a randomly selected bootstrapped sample from the training set. Each tree, thus built is unique in nature and acts as a parallel weak learner. For a given source, each tree votes for it belonging to one of the 8 classes. The fraction of trees out of the entire ensemble voting for a particular class is treated as the class membership probability (CMP) of the given source. For example, if 8 out of 10 trees vote for a particular sample to belong to class AGN, then the sample is said to be AGN with a membership probability of 0.8.
        \subsubsection{Gradient Boosted Decision Tree}
         Gradient Boosted Decision Trees (GBDT) is an ensemble of weak learners \citep{friedman2001greedy}. Compared to a decision tree, where each tree is built independently, in GBDT the trees are built sequentially based on the error of the previous tree. For each newly constructed tree, a loss is calculated based on the error between the predicted and the true values. In the case of the classification algorithm, categorical cross-entropy is used as the loss function and is defined as: 
         \begin{align}
             Loss = -\sum y_i \times log \hat{y_i}
             \label{equ-loss}
         \end{align}
         where $y_i$ is a vector of length 8 (number of classes) with 1 for the true value and 0 otherwise. Here, $\hat{y_i}$ is also a vector of length 8 representing the probability of the object belonging to each class.
         The gradient of this loss function at $(m-1)^{th}$ tree is used to construct a new tree. It is then combined with the previous trees after multiplying it with a weight factor called learning rate $\eta$, which varies from 0 to 1. Essentially, each new tree is built to minimize the error from the previous tree. The main advantage of GBDT over RF is that each newly constructed tree uses the loss from the previous tree and thus tries hard to better classify the previously incorrectly classified sources. The GBDT model can also learn more complex decision boundaries than the RF.

        \subsubsection{Light Gradient Boosted Machine}
       Light Gradient Boosted Machine (LightGBM) was developed by \cite{ke2017lightgbm}. LightGBM is an advanced and efficient version of Gradient Boosted algorithms. Compared to GBDT, where each feature value is compared at the decision nodes of a tree, LightGBM first discretises the value of the input features and then uses these values to construct the decision trees. To make learning more efficient, LightGBM implements two novel techniques, namely, Gradient-based One-Side Sampling (GOSS) and Exclusive Feature Bundling (EFB). With GOSS, LightGBM downsamples the low-gradient examples and upsamples high-gradient examples, which are more difficult to learn. Using EFB, LightGBM bundles the mutually exclusive features (the features that rarely take zero simultaneously) to reduce the dimension of the feature space. Another major capability of LightGBM is that it can handle missing values. It is very useful as we have a large number of missing values in our dataset. It uses the Block Propagation method \citep{josse2019consistency}. In this method for splitting the nodes of the tree, only the available features are used. Wherever a missing value in a sample is found, it is sent to the side that would minimize the final loss.
       
       \begin{table}
        \centering
        \caption{Number of sources under various classes in the training set.}
        \label{tab:train-set}
\begin{tabular}{lcr}
\hline
\textbf{Class}              & \textbf{\% of training set} & \textbf{Number of sources} \\ \hline
AGN                         & 31\%                        & 2395                       \\
STAR                        & 36\%                        & 2790                       \\
YSO                         & 15\%                        & 1149                       \\
HMXB                        & 10\%                        & 748                        \\
LMXB                        & 2\%                         & 143                        \\
ULX                         & 3\%                         & 211                        \\
CV                          & 2\%                         & 166                        \\
PULSAR                      & 1\%                         & 101                        \\
\textbf{Total training set} &                             & \textbf{7703}              \\ \hline
Unidentified Sources        &                             & 269366                     \\
\textbf{Total}              &                             & \textbf{277069}            \\ \hline
\end{tabular}
    \end{table}
    \subsection{Data Normalisation}
    The feature value in the data table varies in order of magnitude. Any ML model, in this scenario, would artificially tend to give more importance to the feature with higher magnitude. Thus it is a general practice to normalise the dataset such that the magnitude variation across features remains uniform before feeding the data to the model. In our case, we normalise the data in such a way that the values lie between 0 and 10 using the following equation,
    \begin{align}
        X_{norm} = 10\times \frac{X-min(X)}{max(X)-min(X)},
    \end{align}
    where $X_{norm}$ is normalised values of the feature $X$, $min(X)$ and $max(X)$ is the minimum and the maximum value of the feature $X$.

    \subsection{Missing value imputation}\label{sec:missing-values}
    In our data table, we compile features from different multi-wavelength catalogues. Due to the difference in coverage of these catalogues and the differences in limiting sensitivity, objects may not be available in all the catalogues. For example, the {\em SDSS} survey coverage is limited to the northern hemisphere. Similarly, we may have missing values in the data table due to differences in the intrinsic luminosity of the source across different wavelengths.
     \begin{figure}
        \centering
        \includegraphics[width=\columnwidth]{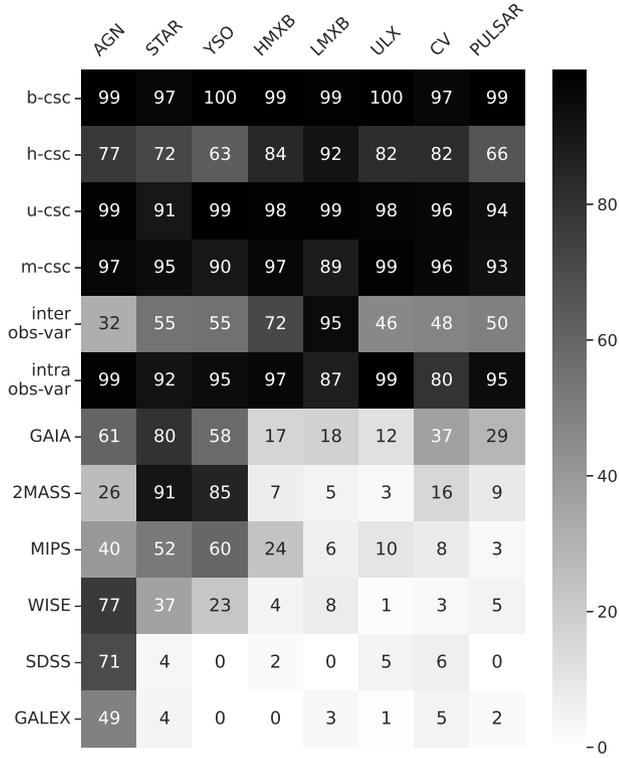}
        \caption{Plot showing the percentage of availability for different features group (see \S \ref{sec:missing-values} for details).}
        \label{fig:available-features}
    \end{figure}
    For example, X-ray binaries in the quiescent stage have lower luminosity in optical-UV and IR but are prominent in X-rays. In the X-ray domain, based on the variability timescales of the objects, the variability features are not available for some of the objects. Figure \ref{fig:available-features} shows the fraction of sources for which the given set of features are available. In the figure, different features from the same instruments are grouped together. For {\em Chandra}, the fluxes in different bands are mentioned separately. The four intra-observation and three inter-observation variability properties are grouped together. We can see that 2MASS, {\em MIPS} and WISE are mostly available only for AGNs, stars and YSOs. The availability of X-ray variability features is significantly higher for X-ray binaries.
    
    Most ML classification models need an input of a fixed size and are incompatible with missing values. One of the methods to avoid missing values is to remove the instances with missing features. We have hardly any source with 100\% features' availability. For RF and GBDT models, we must have a method to handle these missing values. Imputation refers to the method of filling in these missing values with a suitable guess. Imputation can be done using statistics on the available data. In a tabular data format where rows represent the sources and columns represent the features, the missing values can be filled using the mean/mode/median of the feature column.  For RF and GBDT models, we select to impute the missing values using column mode. However, the imputation method may not give satisfactory results for a high percentage of missing values, and in some cases, it is counter-productive to the final output. The RF and GBDT require missing value imputation and, therefore, may not perform well, particularly for minority classes where the percentage of missing values is high. Moreover, in some cases, the missing values themselves may be important features. For example, X-ray binaries (in the quiescent stage) are less likely to be observed in optical wavelengths. Therefore, we have tried to avoid imputation altogether using the model LightGBM as our final classifier. This model is capable of handling the missing values in the data table and also provides better results in every aspect compared to RF and GBDT. 
    
    \subsection{Class imbalance problem}\label{sec:class-imbalance}
         Table \ref{tab:train-set} shows that the numbers of AGNs, stars and YSOs are typically an order of magnitude higher than those of LMXBs, ULXs, CVs and pulsars. Therefore, it is obvious that there is a vast imbalance in the number of training sources, with the majority classes being AGN, star, YSO and the minority classes being LMXB, ULX, CV and pulsar. Any classifier model can achieve higher accuracy by biasing itself towards the majority class and thus would fail to perform on the new data. 
       \begin{figure}
            \centering
            \includegraphics[width=\columnwidth]{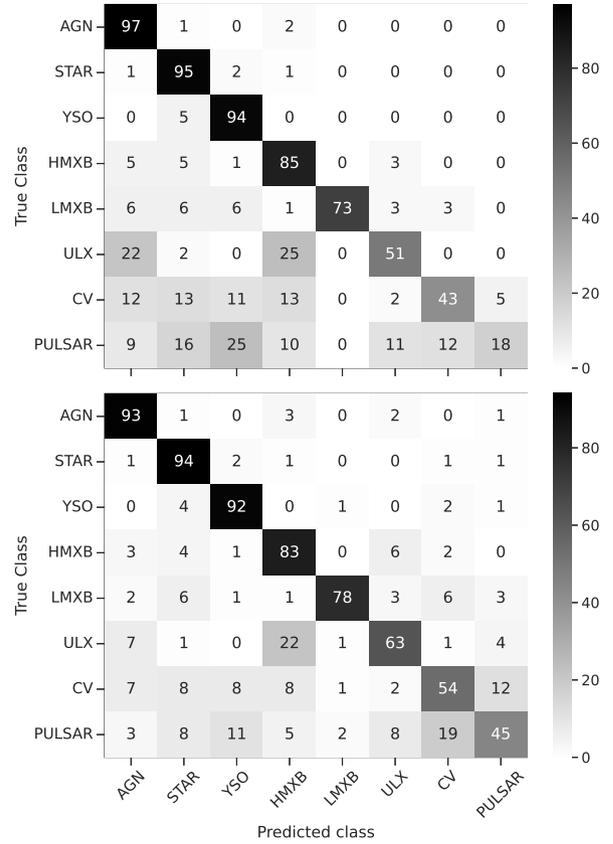}
            \caption{Confusion matrix for Random Forest classifier showing the comparison between no upsampling (top) and upsampling using SMOTE (bottom). The confusion matrix is normalised by the true number of sources available in each class. See \S \ref{sec:class-imbalance} for details.}
            \label{fig:smote-comp}
        \end{figure}
        
        To tackle the class imbalance problem, we use Synthetic Minority Oversampling Technique (SMOTE; \cite{2011arXiv1106.1813C}). In the feature space, it performs linear interpolations between k-neighbouring points (which represent a source in the feature space) and synthetic sources are sampled. Using this technique, each class is sampled such that the number of sources in the minority class becomes equal to the same in the most populous class. To keep our result insensitive to the oversampling, we perform SMOTE only on the training set and not on the validation set. SMOTE is used only for the RF and GBDT models. In LightGBM, we are working with missing data, and SMOTE cannot be performed with missing values. In LightGBM, we use the class-weight technique, which assigns higher weightage to the samples belonging to the minority class in calculating the loss function (Equation \ref{equ-loss}) during training. Essentially in the loss function, we make sure that equal contribution comes from each class.

    \subsection{Strategy for model performance validation}\label{sec:ccv-method}
    We compare the performance of the classifiers using a custom version of $k$-fold cross validation, which we call cumulative $k$-fold cross validation (CCV).
    
    \begin{figure}
        \centering
        \includegraphics[width=\columnwidth]{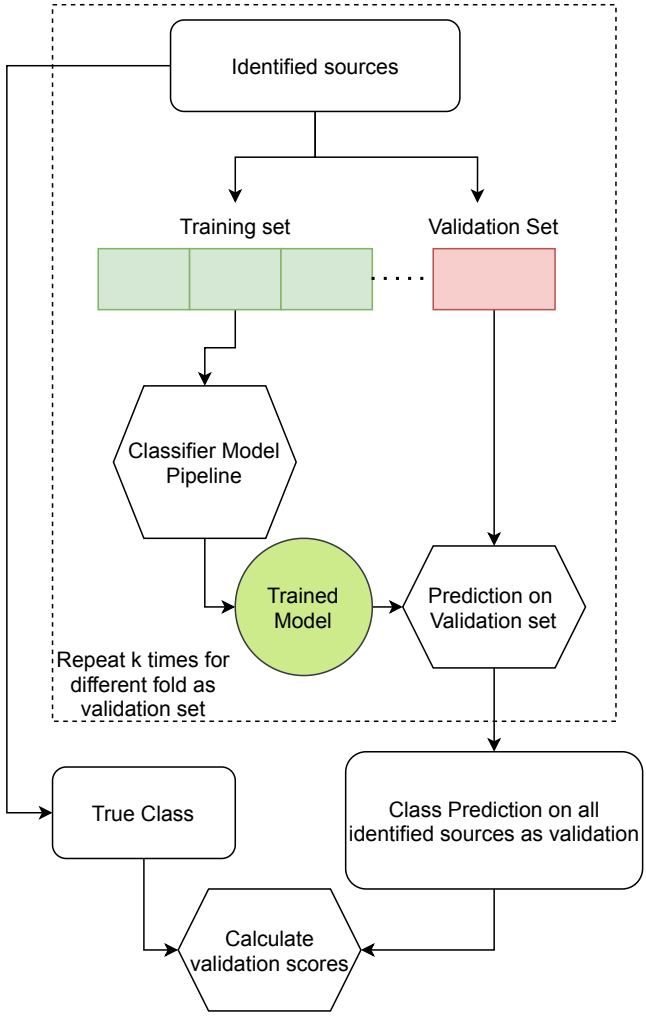}
        \caption{Flowchart showing the cumulative cross validation algorithm. The components inside the dashed box represent one fold of validation. The algorithm is discussed in detail in \S \ref{sec:ccv-method}.}
        \label{fig:ccv-method}
    \end{figure}
     
    The flowchart in the Figure \ref{fig:ccv-method} shows the cumulative cross validation method. Here, we divide the training set into $k$-fold, and we train the classifier using $k-1$ folds and keeping aside $k^{th}$ fold as the validation set in each iteration. After training, we make a prediction on the $k^{th}$ fold (represented by the dashed box in Figure \ref{fig:ccv-method}). The predictions of classes of these samples are then stored in a prediction table. In the next iteration, when a different set of samples are in the validation set, their predictions are stored in the prediction table. In this manner, after $k$ iterations, in the prediction table, we have the predictions for each training source coming only from the iterations when the source was in the validation set. Finally, the elements in the matrix are calculated using this prediction table.
    
    We use precision, recall and F1-score for comparing classifier performance. 
    The precision score is the probability that the predicted class is actually the true class for the sample and is defined as:
    \begin{align}
        precision_A = \frac{TP}{TP+FP},
    \end{align}
    where $precision_A$ represents the precision score for class $A$, $TP$ represents the true positive, i.e., the number predicted as class $A$ actually belonging to class $A$. Here, $FP$ measures the false positive, i.e., the number of samples for which prediction are class $A$, while they belong to some other classes.
    
    In probabilistic terms, the recall score is the probability of identifying the samples truly belonging to that particular class. The recall score for class A is defined as:
    \begin{align}
         recall_A = \frac{TP}{TP+FN},
    \end{align}
    where $FN$ represents the number of samples belonging to class $A$ but predicted to be in a class other than $A$.
    
    F1 score is the harmonic mean of the precision and the recall score: 
    \begin{align}
         F1_A = 2\times \frac{precision_A\times recall_A}{precision_A+recall_A}.
    \end{align}

    We also use Mathew's correlation coefficient (MCC), first introduced by  \cite{MATTHEWS1975442}, which is supposed to be a better representation of classification score, particularly for imbalanced dataset \citep{boughorbel2017optimal}. MCC is defined as 
    \begin{align}
            MCC = \frac{TP\times TN - FP\times FN}{\sqrt{(TP+FP)(TP+FN)(TN+FP)(TN+FN)}}
    \end{align}
    where $TN$ is the number of true negatives. MCC score is comparable to the correlation between the output predictions and the true labels whose values vary between +1, 0 and -1. The perfect prediction means +1, 0 means the model is as good as random predictions, and -1 shows a complete disagreement between the predicted and the true labels.

    We use a confusion matrix for summarising the model performance. An element of the matrix shows the percentage of sources which truly belong to the class given on the y-axis being classified by the model to a class given on the x-axis. The diagonal elements essentially are the recall score for the individual classes. In the CCV, for each iteration, after the dataset is split into training and test set, SMOTE is applied to the training set for the models: RF and GBDT.
    
    Figure \ref{fig:smote-comp} shows the confusion matrix for 20-fold cumulative cross-validation without upsampling (top) and with SMOTE upsampling (bottom) for the Random Forest model. The numbers in the matrix are normalised by the number of sources belonging to the class shown on the vertical axis (true label). The diagonal elements in the figure represent the correct classification, and the rest are wrongly classified. We also notice from the top figure that most of the sources belonging to minority classes, mainly CV and pulsar, are getting classified as AGN, STAR and YSO if no upsampling is done. However, the correctly classified pulsars improve from 18\% to 45\% using SMOTE, and  $\sim 10$\% improvement happens for other minority classes. In general, the reduced contribution at the lower left corner of the bottom figure, as compared to the same in the top figure, shows the effectiveness of SMOTE in reducing the bias of the classifier towards the majority class.
    
    Nevertheless, any class balancing method is not capable of reproducing an equal class distribution. The minority classes: LMXB, CV, ULX, and pulsars, together constitute less than 10\% of the total sample. In such a case, for SMOTE, where we are essentially interpolating source points in the feature space, the number of data points available may not capture the true distribution of the sources and may be affected by the boundary cases. For these minority classes, however, the upsampling shows significant improvement but does not take the score to a satisfactory level.
    
\section{Result and Discussion}\label{sec:result}
    We train the classifier models: RF, GBDT and LightGBM and evaluate the performance using the cumulative $k$-fold cross-validation. We implement the models using \texttt{Scikit-Learn} Python library. For the final selected models, we tune the hyperparameters using FLAML (A Fast Library for Automated Machine Learning \& Tuning) by \cite{2019arXiv191104706W}. To compare the model performance, using the CCV method, we compute the precision, recall and F1 score for each class.

    \begin{table*}
        \centering
        \caption{Precision, recall and F1 score for different classes for RF, GBDT and LightGBM models. The scores are calculated by 20-fold cumulative cross-validation.}
        \label{tab:model-comp-class}
        \begin{tabular}{lccccccccr}
        \hline
        Score$\rightarrow$ & \multicolumn{3}{c}{\textbf{Precision}}  & \multicolumn{3}{c}{\textbf{Recall}}     & \multicolumn{3}{c}{\textbf{F1 score}}  \\
        class$\downarrow$ Model$\rightarrow$ & RF            & GBDT     & LightGBM              & RF           & GBDT         & LightGBM        & RF           & GBDT     & LightGBM      \\ \hline
        AGN                                  & 96.7$\pm$0.1  & 97.8$\pm$0.2 & 96.8$\pm$0.2      & 93.3$\pm$0.2 & 89.6$\pm$0.3 & 97.6$\pm$0.2    & 95.0$\pm$0.1 & 93.5$\pm$0.2 & 97.2$\pm$0.1 \\
        STAR                                 & 95.6$\pm$0.1  & 97.1$\pm$0.2 & 96.0$\pm$0.2      & 94.1$\pm$0.2 & 91.4$\pm$0.2 & 95.7$\pm$0.2    & 95.0$\pm$0.1 & 94.1$\pm$0.1 & 95.9$\pm$0.1 \\
        YSO                                  & 91.6$\pm$0.2  & 91.2$\pm$0.3 & 92.7$\pm$0.3      & 92.4$\pm$0.3 & 93.7$\pm$0.3 & 95.4$\pm$0.3    & 92.0$\pm$0.2 & 92.4$\pm$0.2 & 94.1$\pm$0.2 \\
        HMXB                                 & 79.2$\pm$0.7  & 83.4$\pm$0.8 & 91.6$\pm$0.5      & 83.4$\pm$0.5 & 87.4$\pm$0.7 & 90.7$\pm$0.6    & 81.2$\pm$0.5 & 85.3$\pm$0.5 & 91.2$\pm$0.4 \\
        LMXB                                 & 84.8$\pm$1.3  & 77.4$\pm$2.8 & 94.8$\pm$1.6      & 80.4$\pm$1.1 & 80.8$\pm$0.9 & 80.9$\pm$1.6    & 82.5$\pm$0.8 & 79.1$\pm$1.5 & 87.2$\pm$0.9 \\
        ULX                                  & 52.4$\pm$1.0  & 47.4$\pm$1.1 & 72.2$\pm$1.4      & 63.5$\pm$1.1 & 75.0$\pm$1.1 & 71.1$\pm$1.4    & 57.4$\pm$0.9 & 57.9$\pm$1.0 & 71.5$\pm$1.2 \\
        CV                                   & 49.1$\pm$1.1  & 42.4$\pm$1.0 & 61.5$\pm$1.6      & 53.8$\pm$1.2 & 60.1$\pm$1.6 & 55.3$\pm$1.6    & 51.4$\pm$0.3 & 49.7$\pm$1.1 & 57.4$\pm$1.5 \\
        PULSAR                               & 35.8$\pm$1.9  & 28.3$\pm$1.2 & 42.1$\pm$1.8      & 47.1$\pm$2.6 & 55.8$\pm$1.8 & 44.2$\pm$1.8    & 40.7$\pm$2.1 & 37.6$\pm$1.4 & 43.7$\pm$2.0 \\ \hline
        \end{tabular}
    \end{table*}
    
    We perform several iterations of the 20-fold CCV, each time starting with a new random seed. These iterations are performed to compute the spread in the scoring metrics till the errors stabilize, which occurred after about 15 iterations. For each model, 15 iterations of CCV are performed, and the reported  mean and the standard deviations are estimated over these iterations.
    The precision, recall and F1 score for 20-fold cumulative cross validation are given in Table \ref{tab:model-comp-class}. The performance for majority classes (AGN, star and YSO) with precision, recall and F1 scores greater than 92\% are significantly higher than ULX, CV and pulsar, for which scores are between 40 to 60\%. Performance is moderately good for LMXB and HMXB. We observe a similar trend across all three models. However, when models are compared, LightGBM performs the best for each class, while RF and GBDT have similarly lower performances.  
    
   The LightGBM model performs marginally better than RF for AGN, STAR and YSO, but the scores are significantly higher for LMXB, HMXB and ULX. For minority classes, there is a large difference between precision and recall for RF and GBDT models, whereas these two scores are comparable for LightGBM. For example, GBDT tries more aggressively to increase the prediction chances for the minority class and hence improves the recall score by about 7\% for ULX, CV and pulsars, but this results in a drop in the precision score for the minority classes.
   Similarly, the RF model precision score for pulsars is 35\%, whereas the recall score is 47\%. It means that the model is trying hard to predict more pulsars to achieve a higher recall score, and therefore precision score for pulsars is reduced. This factor is well balanced for LightGBM, resulting highest F1 score across all classes.

    \begin{table}
        \centering
         \caption{MCC score and weighted average precision, recall and F1 score for RF, GBDT and LightGBM models.}
        \label{tab:model-comp}
        \begin{tabular}{lccr}
        \hline
        \multirow{2}{*}{Score} & \multicolumn{3}{c}{Model}                           \\ 
                               & \textbf{RF}     & \textbf{GBDT} & \textbf{LightGBM} \\ \hline
        Precision              & $90.7\pm0.1$   & $91.3\pm0.1$         & $93.2\pm0.1$           \\
        Recall                 & $90.0\pm0.1$   & $89.0\pm0.2$          & $93.2\pm0.1$            \\
        F1 score               & $90.3\pm0.1$   & $89.9\pm0.2$          & $93.2\pm0.1$            \\
        MCC                    & $0.87\pm0.00$ & $0.86\pm0.00$         & $0.91\pm0.00$          \\ \hline
        \end{tabular}
    \end{table}
    Table \ref{tab:model-comp} shows the MCC score and the weighted average of precision, recall and F1 score for RF, GBDT and LightGBM classifier, with the weighting factor being the proportion of sample available in each class. The weighted average is taken to account for the class imbalance. LightGBM has the highest F1 score of 93.3\% whereas the same for RF and GBDT are about 90\%. Once again, it shows that the LightGBM classifier performs the best. The MCC score for LightGBM is the highest (0.91), which means that the prediction made by LightGBM is more correlated with the true value than other models.
    
    \begin{figure}
        \centering
        \includegraphics[width=\columnwidth]{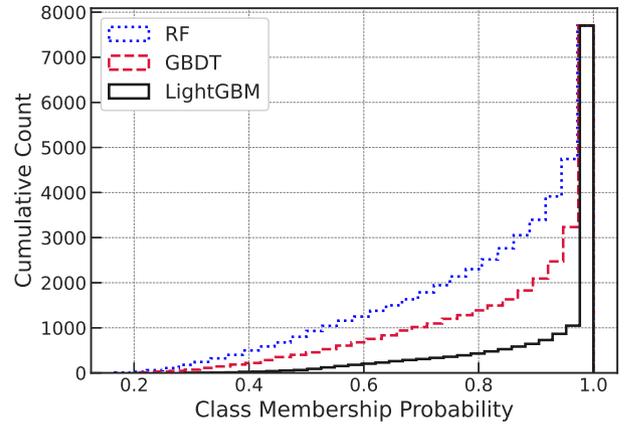}
        \caption{Cumulative histogram of class membership probability for RF, GBDT and LightGBM models, for the training set calculated during cumulative cross validation.}
        \label{fig:cum-prob-dist}
    \end{figure}
    Apart from high precision, recall and F1 score, a good classifier should be capable of predicting the class membership with high confidence. 
    Figure \ref{fig:cum-prob-dist} shows the cumulative histogram of the class membership probabilities (CMP) assigned to the sources in the training set during CCV. The plot shows the histogram for all three models: RF, GBDT and LightGBM. Any point on the plot shows the number of sources with CMP below the value given on the horizontal axis. The more sharply peaked the plot towards unity, the higher the predicted CMP and the more confident the model is. The figure reveals that the LightGBM is the most confidant classifier model among the three, with only about 500 out of 7703 sources below CMP of 0.8, while RF is the least with more than 2000 sources below CMP of 0.8. Therefore, the cumulative cross validation of the models shows that LightGBM is the best classifier model of choice.

\begin{figure}
        \centering
        \includegraphics[width=\columnwidth]{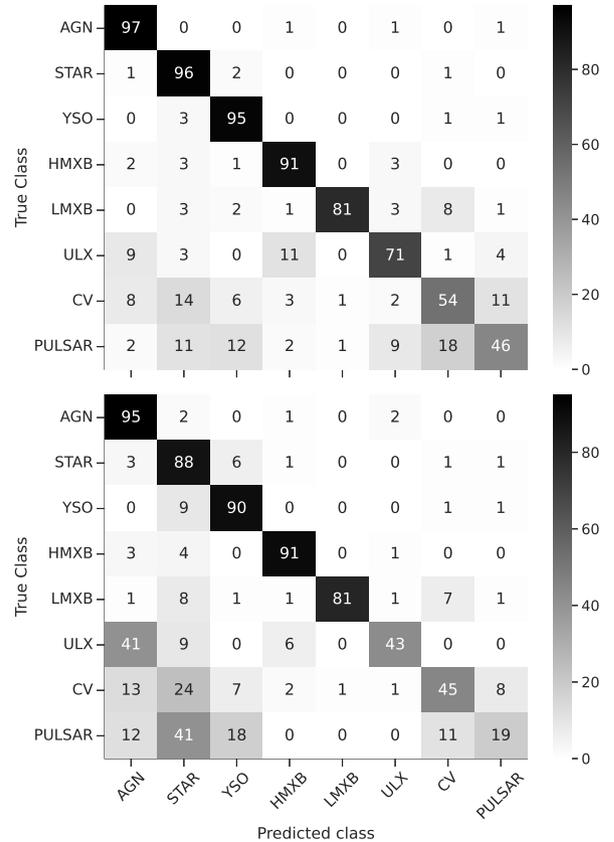}
        \caption{Confusion matrix for the two sample sets: all-features (top) and only X-ray feature (bottom). The vertical axis represents the percentage of sources belonging to a class, and that being classified into various classes are shown on the horizontal axis.}
        \label{fig:confusion-matrix}
    \end{figure}

    To study the importance of multiwavelength features, apart from the base sample, which includes all 41 features mentioned in Table \ref{tab:feature-table}, we create another sample keeping only the X-ray features obtained from CSC 2.0. We perform 15 iterations of the 20-fold CCV for the X-ray sample with the LightGBM model. Then we compare the model performance with the base sample score to see the class-wise dependency on the MW features which are presented in Table \ref{tab:feat-imp-score} and Figure \ref{fig:confusion-matrix}.

    \begin{table}
        \centering
        \caption{Class-wise precision, recall and F1 score for CCV using LightGBM model. The validation is done for various sets of `sample type' indicated in the second column: `all features' where all the 41 features in Table \ref{tab:feature-table} are used and `X-ray', the sample with no optical, UV and IR features.}
        \label{tab:feat-imp-score}
        \begin{tabular}{lcccr}
        \toprule
            \textbf{Class} & \textbf{Sample type}   & \textbf{Precision} & \textbf{Recall} & \textbf{F1 Score} \\
        \midrule
        AGN     & all-features &  96.8$\pm$0.2 &  97.6$\pm$0.2 &  97.2$\pm$0.1 \\
                & X-ray &  91.0$\pm$0.2 &  95.0$\pm$0.2 &  93.0$\pm$0.1 \\
        STAR    & all-features &  96.0$\pm$0.1 &  95.7$\pm$0.1 &  95.8$\pm$0.1 \\
                & X-ray &  89.1$\pm$0.3 &  88.2$\pm$0.2 &  88.6$\pm$0.2 \\
        YSO     & all-features &  92.7$\pm$0.3 &  95.4$\pm$0.3 &  94.0$\pm$0.2 \\
                & X-ray only &  82.9$\pm$0.2 &  89.5$\pm$0.5 &  86.1$\pm$0.3 \\
        HMXB    & all-features &  91.6$\pm$0.5 &  90.5$\pm$0.5 &  91.1$\pm$0.4 \\
                & X-ray &  92.0$\pm$0.5 &  89.9$\pm$0.4 &  90.9$\pm$0.4 \\
        LMXB    & all-features &  94.7$\pm$1.6 &  80.9$\pm$0.7 &  87.3$\pm$0.9 \\
                & X-ray &  95.0$\pm$1.7 &  82.1$\pm$0.5 &  88.0$\pm$0.8 \\
        ULX     & all-features &  72.2$\pm$1.4 &  71.1$\pm$1.5 &  71.6$\pm$1.2 \\
                & X-ray only &  61.3$\pm$2.0 &  42.7$\pm$2.0 &  50.3$\pm$1.9 \\
        CV      & all-features &  61.5$\pm$1.6 &  55.3$\pm$1.7 &  58.2$\pm$1.5 \\
                & X-ray &  56.0$\pm$2.0 &  44.9$\pm$1.8 &  49.8$\pm$1.7 \\
        PULSAR  & all-features &  42.1$\pm$1.8 &  44.2$\pm$2.7 &  43.1$\pm$2.0 \\
                & X-ray &  28.2$\pm$1.9 &  19.0$\pm$1.4 &  22.7$\pm$1.4 \\
        \bottomrule
        \end{tabular}
    \end{table}
    Table \ref{tab:feat-imp-score} shows the mean and standard deviation of the precision, recall and F1 score over 15 times of the 20-fold cumulative cross validation. For each class, the table shows the scores for the two cases: one with X-ray features combined with MW features (all-features) and the other with only X-ray features. We can see a clear performance gain for all the classes when the MW features are used combined with X-ray features. The gain is most pronounced for pulsars with an enhancement of F1 score from 22\% to 43\%. Considering the recall score, 44\% of the pulsars can be retrieved using all features. However, only 19\% of the pulsars are classified properly with only X-ray features. For AGN, if all the MW features are simultaneously removed, the precision, recall and F1 score drop by 5\%, 2\% and 4\%, respectively. Even with only X-ray features, AGNs have a remarkably high F1 score of 93\%. Stars and YSOs, too, follow a similar trend but the dependency on MW features is higher than AGN, with typically 10\% drop in F1 score if MW features are removed. HMXB and LMXB are insensitive to optical/UV and IR features, and there is hardly any drop in performance without MW features. This can be attributed to the fact that X-ray binaries are, in general, faint in optical/UV/IR wavelengths. 
    
    For the ULXs, if MW features are removed, the F1 score drops about 20\%, and it is seen that a large percentage of ULXs are classified as AGNs (Figure \ref{fig:confusion-matrix}, bottom panel). It results in a slight drop in the F1 score for AGN, but it translates to a very high drop in the F1 score of ULX due to a very small population. For CVs, Optical/UV and IR features are important, and the F1 score drops by 10\% when only X-ray features are used.

    \begin{table}
        \centering
        \caption{Same as Table \ref{tab:feat-imp-score}, but the classifier is trained and validated only for the majority class.}
        \label{tab:feat-imp-score-majority-class}
        \begin{tabular}{lcccr}
            \toprule
                 \textbf{Class} & \textbf{Sample type}   & \textbf{Precision} & \textbf{Recall} & \textbf{F1 Score} \\
            \midrule
            AGN & all-features &  98.3$\pm$0.1 &  98.4$\pm$0.1 &  98.3$\pm$0.1 \\
                 & X-ray &  95.5$\pm$0.2 &  96.5$\pm$0.2 &  96.0$\pm$0.1 \\
            STAR & all-features &  97.3$\pm$0.1 &  96.3$\pm$0.1 &  96.8$\pm$0.1 \\
                 & X-ray &  92.7$\pm$0.2 &  89.6$\pm$0.2 &  91.1$\pm$0.2 \\
            YSO & all-features &  93.7$\pm$0.2 &  96.5$\pm$0.2 &  95.1$\pm$0.2 \\
                 & X-ray &  84.4$\pm$0.3 &  91.2$\pm$0.4 &  87.7$\pm$0.3 \\
            HMXB & all-features &  93.9$\pm$0.4 &  93.0$\pm$0.3 &  93.4$\pm$0.2 \\
                 & X-ray &  93.8$\pm$0.4 &  90.7$\pm$0.4 &  92.2$\pm$0.3 \\
            \bottomrule
        \end{tabular}
    \end{table}

     \begin{figure*}
        \centering
        \includegraphics[width=\textwidth]{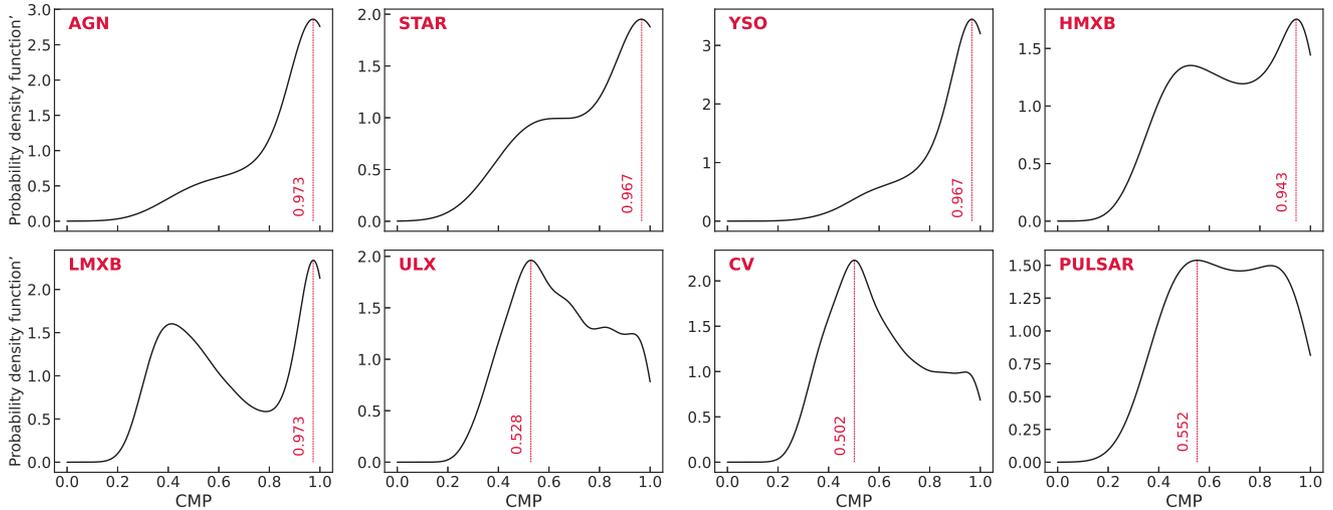}
        \caption{Probability density function (PDF) of the distribution of class membership probability (CMP) of all the unidentified sources predicted using the LightGBM model for different classes (marked on the plots). The most probable value of the class membership probability is shown in the plot with a vertical dashed line.}
        \label{fig:prob_plots}
    \end{figure*}
    Figure \ref{fig:confusion-matrix} shows the confusion matrix for the two cases: one with all the features (top) and the other one with only X-ray features (bottom).  Looking at the ULX row, with MW features, only 9\% (top figure) of ULXs are identified as AGNs, but the same increases to 41\% if only X-ray features are used. Using MW features, the classifier can better separate ULXs from AGNs. Similarly, we see pulsars are most likely to be identified as stars without MW features, and the incorrect prediction increases from 11\% to 41\%. From the confusion matrix, it is evident that the network becomes more biased towards AGN, star and YSO without MW features and the fraction of ULX, CV and pulsar being classified as AGN, star and YSO increases.

    To find the impact of different MW bands, we did a similar analysis adding only optical/UV features with X-ray features, and then only IR features with X-ray features. For AGN and Stars, both Optical/UV and IR features result in identical improvement in the F1 score, but for YSOs addition of IR features has a higher impact than optical/UV features. For ULX and pulsar, each additional MW band (Optical/UV or IR) provide 7-10\% improvement in the F1 scores.

    Even with the multi-wavelength features and class weight balancing, the model does not provide a satisfactory result for the minority classes: LMXB, ULX, CV and pulsars. We attempt to explore to what extent the presence of minority classes affects the classification of majority classes. We train another model using only the four majority classes: AGN, YSO, Stars and HMXB with and without MW features. Table \ref{tab:feat-imp-score-majority-class} shows the scores for the model trained only on the majority classes. Comparing Table \ref{tab:feat-imp-score} and Table \ref{tab:feat-imp-score-majority-class}, the later model has shown a slightly improved performance for these classes. For the all-features case, the F1 scores of AGN, Star and YSO  show $\sim$ 1\%, and HMXB shows $\sim 2\%$ improvement. Considering only the X-ray feature, AGN, Star, YSO and HMXB have 3\%, 3\%, 1\% and 2\% improved F1 scores, respectively. The inclusion of minority classes does not significantly affect the majority class scores if MW features and the X-ray features are used together. We keep the minority classes in our classification scheme; otherwise, these minority class objects will be classified as the majority class.

    From the results discussed above, we select LightGBM as the final classifier with all the 41 features in Table \ref{tab:feature-table} for training and classification of objects into 8 classes.

   The trained LightGBM model is applied to all the 2,69,366 unidentified sources. For each source, the trained model gives a probability of belonging to each of the 8 classes. For a given source, the class having the highest probability is assigned to it, and the corresponding class probability is called the class membership probability (CMP). Thus we get the class identification and the corresponding CMPs for each of the 2,69,366 sources.

    To identify the class-wise confidence of the classifier, we dwell deeper into the corresponding predicted probabilities. After the classification is done, we compute the class-wise probability density function (PDF) of these CMP. For example, 114642 objects out of 269366 are identified as AGNs (Table \ref{tab:new-src}). Using the CMP of these 114642 objects, we calculate the PDF of AGN's CMP. 
    The probability of getting an object of class $A$ with a CMP = $x$, lying between the values $a$ and $b$ is given by the area under the curve of the PDF and is given as
    \begin{align}
        P(a<x<b) = \int_a^b f_A(x)dx,
        \label{eq:prob}
    \end{align}
    where the function $f_A(x)$ is the PDF of the class $A$. It is calculated from the CMP histogram with $N$ bins and counts ($n_i$) in $i^{th}$ bin using the following equation
    \begin{align}
        f_A(x_i) = \frac{n_i}{\sum_0^N {n_i} \times \Delta x},
    \end{align}
    where $\Delta x$ is the bin size. We plot the probability density function in Figure \ref{fig:prob_plots} to present the class-wise distribution of the CMP. The peak of the PDF is marked by red dashed lines, which is the highest CMP value for a given class. The PDF curves show the overall classification confidence for each class. The PDFs are more sharply peaked close to 1.0 for the majority classes (e.g., AGN, STAR, YSO). It means the LightGBM model can predict these classes with high confidence. 

    In contrast, for some other classes (e.g., ULX, CV), the peak appears at lower CMP values, implying a good fraction of sources with lower classification confidence. We find double-peak profiles for some classes (e.g., LMXB). This implies that the nature of some sources of such a class is predicted with low confidence values, while that of other sources of the same class is predicted with high confidence values. The reason could be the availability or unavailability of one or more observational features for certain sources of that class. For example, we find that the inter-observation variability feature is not available for a larger fraction of low-confidence LMXBs.

     \begin{table}
        \centering
        \caption{The number of sources identified in various classes with the LightGBM. The column name `all' represents the sources classified based on maximum CPM of the class.}
        \label{tab:new-src}
        \begin{tabular}{lccr}
        \hline
        \multirow{2}{*}{\textbf{Class}} & \multicolumn{3}{c}{\textbf{Number of sources}}                    \\
                                        & \textbf{all}    & \textbf{CMP$>3\sigma$} & \textbf{CMP$>4\sigma$} \\ \hline
        AGN                             & 114642          & 32600                  & 8574                   \\
        STAR                            & 63967           & 16148                  & 5166                   \\
        YSO                             & 40524           & 5184                   & 208                    \\
        HMXB                            & 8321            & 439                    & 46                     \\
        LMXB                            & 1688            & 197                    & 71                     \\
        ULX                             & 6083            & 50                     & 0                      \\
        CV                              & 10999           & 89                     & 1                      \\
        PULSAR                          & 23142           & 63                     & 0                      \\
        \textbf{Total}                  & \textbf{269366} & \textbf{54770}         & \textbf{14066}         \\ \hline
        \end{tabular}
    \end{table}
    To pick the sources which are identified with a very high membership probability, we set a higher threshold to select only those sources above the CMP threshold.
    Table \ref{tab:new-src} shows the number of newly identified sources in various classes. The number of sources in every class based on the most probable values of CMP (red line in Figure \ref{fig:prob_plots}) is given in the second column (all). Similarly, the number of sources above the probability confidence threshold of $3 \sigma$ (CMP $> 0.997$) and $4 \sigma$ (CMP $> 0.9999$) are shown in the third and fourth column of Table \ref{tab:new-src} respectively. We identify 54,770 new sources in the existing classes with $3 \sigma$ confidence and 14,066 new sources with $4 \sigma$ confidence. This significantly increases the number of sources in various classes.
    
\section{Summary and Conclusions}\label{sec:Summary}

The {\em Chandra} Source Catalogue CSC 2.0 contains $\approx 3,17,000$ sources, including $\approx 2,77,000$ point sources, with a majority of them unidentified. In this work, we implement the decision tree based classifier Light Gradient Boosted Machine to identify the CSC 2.0 objects in the classes of AGN, Star, YSO, HMXB, LMXB, ULX, CV and pulsar.  
For the classification, we use X-ray properties from {\em Chandra}, optical/UV properties from {\em Gaia}, {\em SDSS} and {\em GALEX}, and infrared properties from {\em 2MASS}, {\em WISE} and {\em MIPS-Spitzer}. We train the classifier and applied the trained classier to the unidentified sources to estimate the class membership probabilities of these sources. We achieve a classification weighted precision score of 93\%, recall score of 93\%, F1 score of 93\% and Mathew's Correlation coefficient of 0.91. 

We identify 54,770 new point sources out of which there are 32,600 AGNs, 16,148 stars, 5,184 YSOs, 197 LMXBs, 439 HMXBs, 50 ULXs, 89 CVs and 63 pulsars with a confidence of more than $3\sigma$. Even at a higher confidence (more than $4\sigma$), we get 8,574 AGNs, 5,166 stars, 208 YSOs, 46 HMXBs, 71 LMXBs and 1 CV but not ULXs and pulsars.

The identification of an unknown source as a member of a known class is equivalent to the discovery of a new source of that class.
While the main aim of this paper is to find a suitable classifier and apply it to CSC 2.0 point sources, in a subsequent paper we will  list those  sources which could be assigned to various classes with high significance values and discuss their properties in detail. Finally, we believe that our method can be reliable and promising for other catalogues as well.

Very recently, we came across the work by \cite{2022ApJ...941..104Y} which
aims to develop a pipeline called MUWCLASS using a multiwavelength approach to classify sources in CSC 2.0. They applied the pipeline to a sample of 66369 CSC 2.0 sources. The training dataset, peer-reviewed catalogues, the choice of
classifier, the validation methodology and the overall list of
unidentified sources make our work unique.

\section*{Acknowledgements}

This research has used the {\em Chandra} Data Archive and the {\em Chandra} Source Catalogue, and software provided by the {\em Chandra} X-ray Center (CXC) in the application packages CIAO and Sherpa; NASA/IPAC Extragalactic Database (NED), which is operated by the Jet Propulsion Laboratory, California Institute of Technology, under contract with the National Aeronautics and Space Administration; data from the European Space Agency (ESA) mission {\it Gaia} (\url{https://www.cosmos.esa.int/gaia}), processed by the {\it Gaia} Data Processing and Analysis Consortium (DPAC,\url{https://www.cosmos.esa.int/web/gaia/dpac/consortium}); the cross-match service provided by CDS, Strasbourg; Multiwavelength visualisation tool by \cite{2021RNAAS...5..102Y} to identify the colour-colour properties to be used.

This publication makes use of data products from the Wide-field Infrared Survey Explorer, which is a joint project of the University of California, Los Angeles, and the Jet Propulsion Laboratory/California Institute of Technology, and NEOWISE, which is a project of the Jet Propulsion Laboratory/California Institute of Technology. WISE and NEOWISE are funded by the National Aeronautics and Space Administration.
\section*{Data Availability}

All the data accumulated/generated in this work will be made available on a publicly accessible portal which is in the development phase. The portal will include the multi-wavelength data for the 2,77,069 sources. The portal will also allow the user to perform a cone search based on the coordinates. The classification table and the class membership probabilities generated in this work will be integrated into the portal such that the user will be able to filter the search for their class of interest with their selected confidence threshold. The sources identified with very high confidence will be released as a machine-readable data table in a subsequent paper. The training dataset with the corresponding reference will be shared as a CSV table at the reader's request. 


\bibliographystyle{mnras}
\bibliography{references} 




\appendix


\bsp	
\label{lastpage}
\end{document}